\documentclass[showpacs,preprintnumbers,amsmath,amssymb]{revtex4}
\usepackage{graphicx}
\usepackage{epsf}

\begin{document}

\title{Influence of high-angle grain boundaries on the charge order
formation in the $\mathrm{YBa_2Cu_3O_{7-\delta}}$.}
\author{Vladimir S. Boyko and Roman Ya. Kezerashvili}
\affiliation{\mbox{Physics Department, New York City College
of Technology, The City University of New York,} \\
Brooklyn, NY 11201, USA \\}

\begin{abstract}
 We are examining the possibility of the formation of charge order in the
high-temperature superconductor $\mathrm{YBa_2Cu_3O_{7-\delta}}$ (YBCO) due
to interaction between the charged oxygen vacancies or di-vacancies. The
molecular dynamics method is used to analyze the displacement fields around
these defects. The distribution of displacements around a single charged
oxygen vacancy and di-vacancy, determination of binding energy of oxygen
vacancy in di-vacancy demonstrate that there is, in principle, the
possibility of the charge order formation in YBCO by charged oxygen
vacancies or di-vacancies. It is shown that the charge order formation first
of all should be formed near crystal lattice defects and that the high-angle
grain boundaries (GBs) regions are preferable places for this formation. The
adsorption capability of high-angle GBs with respect to the stripe embryo
formation is determined. It is shown that there is a proportional dependence
between the repetition distance along the high-angle GBs and an energy
advantage of the stripe embryo formation in GBs.

\end{abstract}

\maketitle

%\thispagestyle{empty}

%-----------------------------------------------------------------------------------------------------------------------
%-----------------------------------------------------------------------------------------------------------------------

\section{Introduction}
\label{intro}

There are many circumstances in which the charge order plays a significant
role in the physics of electronically interesting materials \cite%
{RobertsonKivelsonFradkinFungKapitulnikPhysRevB2006}. The 1D charge order
intensively discussed now in the physics of high-temperature
superconductivity (see, for example, \cite%
{RobertsonKivelsonFradkinFungKapitulnikPhysRevB2006,KivelsonBindlossFradkinOganesyanTranquadaKapitulnikHowaldRevModPhys2003,ZaanenNature2006}%
). In this case, a spin and charge separate into different linear regions or
stripes \cite%
{ZaanenGunnarsonPhysRevB89,MachidaPhysicaC89,EmeryKivelsonLinPhysRevLett90}.
It was discovered in \cite{TranquadaSternliebAxeNakamuraUchidaNature95} that
small changes in the crystal structure of high-$T_c$ superconductors can
cause disappearance of superconductivity with a static stripe phase taking
over. According to Ref. \cite{ZaanenNature2006}, the existence of static
stripes, initially contentious, is now generally accepted. However, the
structure of charge stripes and conditions of their formation are not well
understood.

The consideration \cite%
{KhomskiiKugelEurophysLett2001,KhomskiiKugelPhysRevB2003} showed, that the
sheet or stripe phases appears quite naturally if the dominant interaction
is the elastic interaction between impurities which is a long--range and
intrinsically anisotropic (attractive in certain directions and repulsive in
others). It is reasonable to suggest that, in the case of the high-$T_c$
superconductors, more perspective is the consideration of a long--range
elastic interaction between oxygen vacancies as the factor stabilizing
charge order. Usually, the oxygen vacancies are much more abundant in the
high-temperature superconductors than impurities because of the
stoichiometry reasons. To perform the scheme considered in Ref. \cite%
{KhomskiiKugelEurophysLett2001} we need to know the ``strength'' of the
point defect -- a local change of the lattice volume at the vacancy
location. Then, following the method developed in Ref. \cite%
{EshelbySolidStatePhysics56}, we should calculate the elastic energy of
interaction of two vacancies (the elastic problem that has no explicit
solution for the orthorhombic lattice). We propose to examine the
possibility of the formation of charge order due to interaction between the
charge oxygen vacancies using computer simulation by molecular dynamics
method. The possibility of the formation of charge order in the classic
high-temperature superconductor YBCO due to interaction between the charged
oxygen vacancies or di-vacancies is examined. It was interesting to check
this possibility also because the evidences were found of the connection
between the charge distribution and oxygen defects in the YBCO. The
discovery of oxygen "superstructures" in cuprate materials \cite%
{StrempferZegkinoglouRuttZimmermannBernhardLinWolfKeimerPhysRevLett2004,IslamLiuSinhaLangMossHaskelSrajerWochnerLeeHaeffnerWelpPhysRevLett2004}
could help to shed light on the problem. In Ref. \cite%
{StrempferZegkinoglouRuttZimmermannBernhardLinWolfKeimerPhysRevLett2004},
authors observed an ordered superstructure with a periodicity of four unit
cells in materials that contained oxygen vacancies, but not in samples that
did not contain the oxygen defects. The formation of the superstructure
depended on the oxygen concentration. In similar experiments \cite%
{IslamLiuSinhaLangMossHaskelSrajerWochnerLeeHaeffnerWelpPhysRevLett2004},
"nanodomains" of displaced copper, barium and oxygen atoms in the YBCO
crystals were observed. The presence of these domains indicates that an
ordered pattern of oxygen vacancies forms, leading to the same
superstructure seen by \cite%
{StrempferZegkinoglouRuttZimmermannBernhardLinWolfKeimerPhysRevLett2004}. As
it was noted in \cite%
{IslamLiuSinhaLangMossHaskelSrajerWochnerLeeHaeffnerWelpPhysRevLett2004},
the oxygen vacancies in the Cu-O chains tend to form superstructures
according to the scheme proposed by de Fontaine and co-authors \cite%
{CederAstaCarterKraitchmandeFontaineMannSluiterPhysRevB1990}. In the paper 
\cite{deFontaineOzolinsIslamMossPhysRevB2005}, de Fontaine and co-authors
present electronic structure calculations performed prior to the
experimental observations \cite%
{IslamLiuSinhaLangMossHaskelSrajerWochnerLeeHaeffnerWelpPhysRevLett2004}
which prove that observed diffraction patterns are due to static atomic
displacements around missing rows of oxygen atoms in the Cu-O plane. The
phenomenon of ordering is named in \cite%
{deFontaineOzolinsIslamMossPhysRevB2005} as O-compositional stripes.

Even though, according to \cite%
{LeBoefDoiron-LeyraudLevalloisDaouBonnemaisonHusseyBalicasRamshawLiangBonnHardyAdachiProustTailleferNature2007}%
, there is no so far unambiguous direct evidence for a static spin/charge
density-wave order in the YBCO because the putative density-wave phase in
the YBCO involves fluctuating rather than a static order \cite%
{KivelsonBindlossFradkinOganesyanTranquadaKapitulnikHowaldRevModPhys2003},
stripe-like regions were used in the atomic consideration of the structure
of fluctuating density wave in the YBCO \cite{PfleidererHacklNature2007}.
This consideration applied recently by \cite{PfleidererHacklNature2007} for
explanation of experimental observations \cite%
{LeBoefDoiron-LeyraudLevalloisDaouBonnemaisonHusseyBalicasRamshawLiangBonnHardyAdachiProustTailleferNature2007}
of electron pockets in the Fermi surface of the hole-doped YBCO. Actually,
in the Fig. 1 of \cite{PfleidererHacklNature2007}, we can see the positively
charged holes situated at the oxygen di-vacancies organized in the
stripe-like regions.

%-----------------------------------------------------------------------------------------------------------------------
%-----------------------------------------------------------------------------------------------------------------------

\section{Method}
\label{sd}

As the first step we analyzed the displacement field around a single charged
oxygen vacancy in the bulk. We determined the spatial distribution of
binding energy of charge vacancies in the di-vacancy and the displacement
field around the di-vacancy. Finally we considered the conditions of the
charge order formation in the vicinity of the high-angle GBs in the YBCO. It
is noticed by Pennycook and co-workers \cite%
{Klie1BubanVarelaFranceschettiJoossZhuBrowningPantelidesPennycookNature2005}
that first-principles calculations for realistic YBCO grain-boundary
structures are computationally prohibitive. We used the molecular dynamics
method in this work because it allowed us to consider consistently both bulk
and defect regions of the crystal lattice. The molecular dynamics methods
exploited in calculations are described in Refs. \cite%
{BoykoLevinePhysRevB2001} and \cite{BoykoKezerashviliLevinePhysRevB2004}.
The substantial feature of the present work is the computer simulation of
point defects in models of the ideal crystal lattice of the YBCO. It is
worth mentioning that the computer simulation of point defects in the YBCO
was done in a number of articles (see, for example, Ref. \cite%
{ZhangCatlowMolecularSimulation94}), especially for calculations of the
diffusion of the single oxygen vacancy in the perfect crystal lattice. In
our case, we create also simultaneously two or three point defects (charged
oxygen vacancies). There are four different positions of oxygen atoms in the
crystal lattice of the YBCO. The vacancy was created at the position of O4
in the layer Cu1 - O4 (the layer representation \cite%
{PooleFarachCreswikSuperconducitvity95} of the YBCO lattice cell was used).
The configuration of the models is as follows: the $X$-axis is directed
along $a$-axis of crystal lattice, the $Y$-axis -- along $b$-axis, and the $Z
$-axis -- along $c$-axis of the lattice, and the $\mathrm{XOY}$-plane
coincides with the basal $ab$-plane of the lattice. The creation of the
charged vacancy is equivalent to the appearance of the charge $q = 1.3$ $e$,
where $e$ is the elementary charge.

In this work, we are considering also the creation of charged oxygen
vacancies in the most perfect large-angle GBs -- the coincident site lattice
GBs with small $\Sigma$ ($\Sigma$ is inverse of the density of coincident
sites if lattices of neighboring grains are assumed to fill all space). We
will characterize the GBs by the direction of the axis of misorientation,
the angle of misorientation relative to this axis, the geometrical plane of
the GB which is chosen coinciding with some simple crystal lattice plane,
and the repetition distance $R_{d}$ along the GB. We consider the
symmetrical large-angle tilt GBs with the misorientation axes [001]: $\Sigma5
$ $(310)_1/(310)_2$, misorientation angle $\theta = 36.87 \, \mathrm{^{\circ}%
} $, repetition distance along the GB $R_d$ = 3.164$a$, where $a$ is the
lattice constant along $a$-axis ($a = 0.380 \, \mathrm{nm} $); $\Sigma5$ $%
(210)_1/(210)_2$, $\theta = 53.13 \, \mathrm{^{\circ}} $, $R_d$ = 2.238$a$; $%
\Sigma13$ $(510)_1/(510)_2$, $\theta = 22.62 \, \mathrm{^{\circ}} $, $R_d$ =
5.100$a$; $\Sigma13$ $(320)_1/(320)_2$, $\theta = 67.38 \, \mathrm{^{\circ}} 
$, $R_d$ = 3.537$a$; $\Sigma17$ $(410)_1/(410)_2$, $\theta = 28.07 \, 
\mathrm{^{\circ}} $, $R_d$ = 4.124$a$; $\Sigma17$ $(530)_1/(530)_2$, $\theta
= 61.93 \, \mathrm{^{\circ}} $, $R_d$ = 5.839$a$. Here, the boundary planes
common for the two grains are given in parenthesis, and the indexes 1 and 2
refer to the two neighboring crystals of the bi-crystal. The GBs $\Sigma 5$, 
$\Sigma 13$, and $\Sigma17$ were chosen as the most perfect tilt GBs that
cover a wide range of misorientation angles and are experimentally
observable in the YBCO. In the GBs simulation, the $\mathrm{XOZ}$-plane
coincides with the geometrical plane of the GB. The $\mathrm{X}$-axis lies
in the geometrical plane of the GB. The $\mathrm{Y}$-axis is perpendicular
to the geometrical planes of the GB. The $\mathrm{Z}$-axis is directed along
the c-axis of the crystal lattice and coincides with the misorientation
axis. The extent of the model along the $\mathrm{X}$-axis is equal in all
cases to one period of the coincident-site-lattice of the corresponding GB.
The total number of ions in each model is around 600.

In Ref. \cite{ChaplotPhysRevB90}, the interatomic potential for the YBCO was
proposed. It includes pair-wise interactions consisting of the four parts:
the Coulomb potential (the relative values of the charges on different atoms
are determined from the requirement of producing a stable and reasonable
crystal structure); the Born-Mayer-type short-range repulsive potential; the
van der Waals-type potential; the Lippincott--Schroeder-type covalent
potential. This interatomic potential for the YBCO has proved successfully
in calculations of the minimum energy structure, the bulk modulus, the
phonon spectrum, and the orthorhombic-to-tetragonal phase transition. In our
study, we employ the potential \cite{ChaplotPhysRevB90}, as well as the
Ewald method of calculation. Periodic boundary conditions were applied at
all the outer faces of the computational cell. The corresponding classical
equations of motion were solved using the velocity form of the Verlet
algorithm. Verlet algorithms is one of the most common ``drift-free''
higher-order algorithms \cite{GouldTobochnikComputerSimulationMethods88}.
The initial time step in all calculations was not greater than $2.5 \cdot
10^{-15} \, \mathrm{s} $. In each case of the simulation, we started with an
initial configuration of the model in which all atoms, with exception of
atoms removed for vacancies formation, are in their position when vacancies
were absent. It took usually several thousands time steps to achieve the
relaxed configuration. The comparison of the initial and final (relaxed)
configurations allowed us to determine the atomic displacement vector for
each atom. These displacements are shown in Fig. 1 and Fig. 2 as arrows.

\begin{figure}[tbp]
\includegraphics[width = 3.0in] {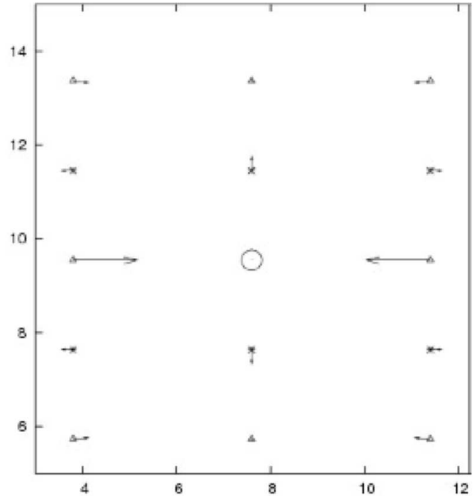}
\caption{Atomic displacements (shown as arrows) around a single
charged vacancy O4 in the crystal lattice of the YBCO. The vacancy is
situated in the basal plane that is parallel to the XOY plane of the model
(the part of its top view). Only the displacements of atoms situated at the
same basal plane are shown. The magnification of arrows is 4. All distances
are given in angstroms. Designations of different types of atoms: Cu1 - $%
\divideontimes$, O4 - $\triangle$.} 
%Y - +, Ba - $\times$, Cu1 - $\divideontimes$, Cu2 - $\boxdot$, O1 -  $\blacksquare$, O2 - $\odot$, O3 - $\bullet$, O4 - $ \triangle$.}
\label{crystal1}
\end{figure}

\begin{figure}[tbp]
\includegraphics[width = 3.0in] {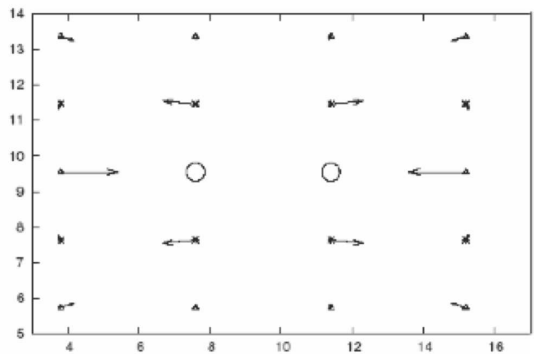}
\caption{ Atomic displacements (shown as arrows) around a charged
di-vacancy O4 in the crystal lattice of the YBCO. The di-vacancy is formed
by two charged vacancies O4 lying at the basal plane. They are the
first-neighbor vacancies in the direction of $a$-axis ($X$-axis in our
consideration). The distance between these two vacancies is the lattice
constant $a$. Only atoms situated at the basal plane and the components of
their displacements in this plane are shown. The magnification of arrows is
2. All distances are given in angstroms. Designations of different types of
atoms: Cu1 - $\divideontimes$, O4 - $\triangle$.} 
%Y - +, Ba - $\times$, Cu1 - $\divideontimes$, Cu2 - $\boxdot$, O1 -  $\blacksquare$, O2 - $\odot$, O3 - $\bullet$, O4 - $ \triangle$.}
\label{crystal2}
\end{figure}
%-----------------------------------------------------------------------------------------------------------------------
%-----------------------------------------------------------------------------------------------------------------------

\section{Results and Discussion}
\label{disc}

The results of our calculations for the atomic displacements around a single
charged vacancy O4 and a charge di-vacancy O4 in the crystal lattice of the
YBCO are shown in Fig. 1 and Fig. 2. The analysis of the distribution of the
atomic displacements around the single charged oxygen vacancy in ideal
crystal lattice showed that this strain field is comparatively long-ranged:
substantial displacements are demonstrated by atoms located at the positions
of the 8th neighbors. The displacement field created by the presence of the
charged oxygen vacancy is anisotropic. There are atoms that are displaced in
the direction of the vacancy and there are atoms that are displaced in the
opposite direction. It is clearly seen from Fig. 1: atoms O4 situated on the
distances $r = a$ from the vacancy in the basal Cu -- O plane are displaced
in the direction of the vacancy with the magnitude of the displacement
vector $\Delta r = 0.089 a$; atoms Cu1 situated on the distances $r = 0.503 a
$ are displaced from the vacancy with the magnitude of the displacement
vector $\Delta r = 0.020 a$. The same type of displacements are occurred for
atoms outside of the basal plane: atoms O1 situated on the distances $r =
0.693 a$ from vacancy are displaced in the direction of the vacancy with the
magnitude of the displacement vector $\Delta r = 0.062 a$; atoms Ba situated
on the distances $r = 0.767 a$ are displaced from the vacancy with the
magnitude of the displacement vector $\Delta r = 0.021 a$.

Actually these results mean that some regions around the vacancy are in the
state of compression, others are in the state of tension. When regions in
the state of compression of one vacancy will be overlapped with regions in
the state of tension of another vacancy, the total elastic energy will
decrease. When like regions of two vacancies will be overlapped, the total
elastic energy will increase. This should lead to the appearance of the
elastic interaction between vacancies. This interaction is attractive at
some mutual orientations and repulsive at others. On the atomic level, it
should be revealed in a formation of stable di-vacancies with the positive
binding energy of the vacancy in the di-vacancy at some mutual orientations
of two vacancies in the di-vacancy. This assumption is supported by
determination of the binding energy $E_b^{V+V}$ of a charged oxygen vacancy
in the di-vacancy and illustrated in the Table 1. The determination was
performed as follows. According to our data, the energy of formation of the
single charge oxygen vacancy O4 in the crystal lattice of YBCO equals $E^{V}
= 0.823 \, \mathrm{eV} $. (The calculations of O4 vacancy formation energies
in the strained and unstrained crystal lattice of YBCO by using
density-functional theory in the local-density approximation yielded
according to Ref. \cite%
{Klie1BubanVarelaFranceschettiJoossZhuBrowningPantelidesPennycookNature2005}
correspondingly $E^{V} = 1.40 \, \mathrm{eV} $ and $E^{V} = 1.90 \, \mathrm{%
eV} $.) The formation energy of the di-vacancy in the bulk from the two
first-neighbor vacancies in the direction of the $a$-axis according to our
data equals $E^{V+V} = 1.504 \, \mathrm{eV} $. Then the binding energy $%
E_b^{V+V}$ of a charged oxygen vacancy in the di-vacancy equals $E_b^{V+V} =
2E^{V} - E^{V+V} = 0.142 \, \mathrm{eV} $. It is positive. Another
configuration of two vacancies in the di-vacancy with positive binding
energy is for the second-neighbor vacancies in the direction of the $b$-axis
($E_b^{V+V} = 0.038 \, \mathrm{eV} $). Other configurations of two vacancies
in the di-vacancy have negative binding energies.

\begin{table}[htbp]
\begin{center}
\leavevmode
\begin{tabular}{|p{1.8cm}|p{3cm}|p{3cm}|p{2cm}|}
\hline
Neighbor & $d_v\hspace{.5cm}(\text{a,b})$ & $d_v\hspace{.5cm}(\text{\AA })$
& $E^{V+V}_b\hspace{.3cm}(\text{eV})$ \\ \hline
$1st$ & a & $3.800$ & $+ 0.142$ \\ \hline
$2nd$ & b & 3.820 & $- 0.699$ \\ \hline
$3rd$ & $\sqrt{a^2 + b^2}$ & 5.388 & $- 0.192$ \\ \hline
$4th$ & 2a & 7.600 & $- 0.295$ \\ \hline
$5th$ & 2b & 7.640 & $+ 0.038$ \\ \hline
$6th$ & $\sqrt{a^2 + (2b)^2}$ & 8.533 & $- 0.031$ \\ \hline
\end{tabular}
\end{center}
\caption{Binding energy $E^{V+V}_b$ of the charged O4 vacancy in the
di-vacancy depending on a distance $d_v$ between vacancies in Cu1 -- O4
plane of the high-temperature superconductor $\mathrm{YBa_2Cu_3O_{7-\protect%
\delta}}$.}
\label{tab:twin}
\end{table}

% \newpage

\begin{table}[htbp]
\begin{center}
\leavevmode
\begin{tabular}{|p{3.0cm}|p{1.5cm}|p{1.5cm}|p{1.5cm}|p{1.5cm}|p{1.5cm}|p{1.5cm}|}
\hline
$R_d (a)$ & $2.238$ & $3.164$ & $3.537$ & $4.124$ & $5.100$ & $5.839$ \\ 
\hline
$GB $ & $\Sigma5$ & $\Sigma5$ & $\Sigma13$ & $\Sigma17$ & $\Sigma13$ & $%
\Sigma17$ \\ \hline
$\theta$ & $53.13 \, \mathrm{^{\circ}} $ & $36.87 \, \mathrm{^{\circ}} $ & $%
67.38 \, \mathrm{^{\circ}} $ & $28.07 \, \mathrm{^{\circ}} $ & $22.62 \, 
\mathrm{^{\circ}} $ & $61.93 \, \mathrm{^{\circ}} $ \\ \hline
$E_b^{GB+emb} (eV)$ & $-6.51$ & $10.13$ & $38.06$ & $34.08$ & $51.13$ & $%
107.77$ \\ \hline
\end{tabular}
\end{center}
\par
\caption{Adsorption capability of grain boundaries in the $\mathrm{%
YBa_2Cu_3O_{7-\protect\delta}}$ with respect to the stripe embryo formation.}
\label{tab:adsorption}
\end{table}

The distribution of displacements around the di-vacancy is shown in Fig. 2.
Di-vacancy is formed by two charged vacancies O4 situated at the distance of
one lattice constant $a$ from each other in the basal plane. In Fig. 2, only
the displacements of atoms lying in the same basal plane are shown.
Therefore, there is a good possibility to compare the displacements fields
around the single vacancy and di-vacancy. It should be pointed out that the
magnification of the displacements in Fig. 1 is 4, but in Fig. 2 it is only
2. There are following displacements around the di-vacancy: atom O4 is
displaced in the direction of the di-vacancy with the magnitude of the
displacement vector $\Delta r = 0.212a$, which is 2.390 times greater than
for a single vacancy; atom Cu1 is displaced from the di-vacancy with the
magnitude of the displacement vector $\Delta r = 0.105a$, which is more than
5 times greater than for a single vacancy. Therefore, based on these results
and by analogy with displacement field of the single vacancy, one can expect
that anisotropy of the displacement field around di-vacancies should lead to
the appearance of interaction between them: the attraction at some mutual
orientations and repulsion at others. Some specific positions of the
di-vacancies should result in decrease of the elastic energy and lead to a
formation of stable configurations of the di-vacancies with the positive
binding energy of the vacancies in such configurations. The consideration 
\cite{KhomskiiKugelEurophysLett2001,KhomskiiKugelPhysRevB2003} showed, that
if the interaction of the impurities in crystals is long-range and
intrinsically anisotropic (attractive in certain directions and repulsive in
others), then the charge order may appear. It was shown above that
interaction between charged oxygen vacancies and di-vacancies in the YBCO is
long-range and intrinsically anisotropic (attractive in certain directions
and repulsive in others). Thus, we can expect that there is, in principle,
the possibility of the charge order formation in the YBCO by vacancies or by
di-vacancies. But actually important question is what should be the trend
when more charge vacancies will be considered? To get the answer on this
question, we examined the energy formation of the small chain of three
charged first neighbor oxygen vacancies along the $a$-axis.

The formation energy of the small chain of three charged first neighbor
oxygen vacancies along the $a$-axis according to our data equals $E^{V+V+V}
= 2.939 \, \mathrm{eV} $. Therefore, the formation energy of additional
third vacancy equals $1.435 \, \mathrm{eV} $. It is greater than $E^{V}$. As
a result, the binding energy of the vacancy in this configuration is
negative and equals $-0.470 \, \mathrm{eV} $. Thus, there is the energy
disadvantage of the formation of charge order by comparatively long chains
of charged vacancies. Therefore, we can expect that the charge order
formation by charged oxygen vacancies first of all will be formed near the
crystal lattice defects. Most relevant defects are the GBs. We can expect
the excess of oxygen vacancies both in the low-angle and high-angle GBs
(see, for example, \cite%
{Klie1BubanVarelaFranceschettiJoossZhuBrowningPantelidesPennycookNature2005,BoykoKezerashviliLevinePhysRevB2004}%
).

It was found in \cite{AndoSegavaKomiyaLavrovPhysRevLett2002} that the
in-plane resistivity anisotropy in single crystals of YBCO gives evidence
for conducting charge stripes. The data indicate that intrinsically
conducting stripes govern the transport in samples with $T_c$ of as high as $%
50 \, \mathrm{K} $. It is well known that GBs play extremely important role
in transport properties of the YBCO polycrystals (see, for example, the
review \cite{HilgencampMannhartRevModPhysics2002}). Therefore, it is
important to understand relationship between charge stripes formation, their
transport properties and the GBs in the YBCO. It is logical to suggest that
transparency or non-transparency of GBs with the respect to the stripes
depends on the energy advantage or disadvantage of stripe embryo formation
in the GBs.

\begin{figure}[tbp]
\includegraphics[width = 2.0in] {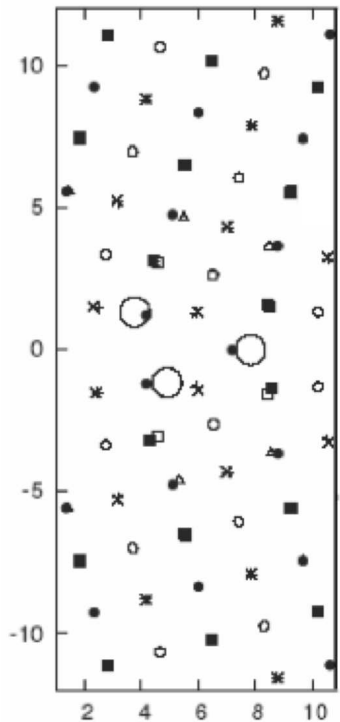}
\caption{ Projection of the positions of atoms at the grain boundary $%
\Sigma17$, $\theta = 28.07 \, \mathrm{^{\circ}} $ with the stripe embryo
inside (consisted of three vacancies) on the plane XOY (this plane coincides
with the basal plane of the YBCO crystal lattice). All distances are given
in angstroms. Designations of different types of atoms: Y - +, Ba - $\times$%
, Cu1 - $\divideontimes$, Cu2 - $\boxdot$, O1 - $\blacksquare$, O2 - $\odot$%
, O3 - $\bullet$, O4 - $\triangle$.}
\label{crystal3}
\end{figure}

\begin{figure}[tbp]
\includegraphics[width = 3.0in] {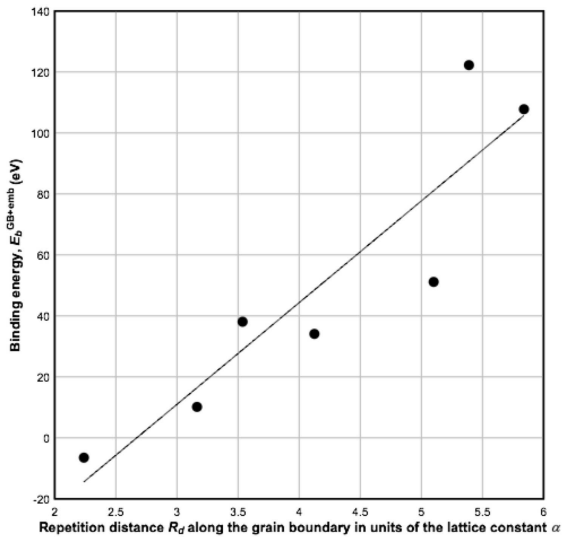}
\caption{ Binding energy stripe embryo with grain boundary versus
repetition distance along grain boundary.}
\label{crystal4}
\end{figure}

We consider as a model of the stripe embryo configuration of three charged
oxygen vacancies in which one vacancy is located directly in the GB plane,
two other vacancies are at first neighbors positions of O4 atoms. They are
situated from both sides of the GB plane and each vacancy in its own grain
as shown in Fig. 3. Generally speaking, we can imagine other configurations
of the stripe embryo, for example, the stripe embryo as a linear-type
configuration of the vacancies piercing the GB. But, in the case of the
linear-type configuration, the stripes in the neighboring grains should
start from different points of the GB. So, we can expect smaller their
influence on the transport properties of the GB. The projection of the
positions of atoms at the GB $\Sigma 17$, $\theta = 28.07^\circ$ with the
stripe embryo inside on the plane XOY is shown in Fig. 3.

To determine the adsorption capability of the large-angle GBs with respect
to the embryo of stripes formation we expanded approach \cite%
{BoykoKezerashviliLevinePhysRevB2004}, where the computer simulation of
single vacancies in the YBCO was done. The adsorption capability of the GB
can be characterized by $E_b^{GB+emb}$ -- the binding energy of the stripe
embryo with the GB. $E_b^{GB+emb}$ can be estimated as follows: 
\begin{equation}  \label{eq:ratio}
E_b^{GB+emb} = E^{GB} - E^{GB+emb},
\end{equation}

where $E^{GB}$ is the energy of the microcrystallite containing the perfect
grain boundary and $E^{GB+emb}$ is the energy of the microcrystallite
containing grain boundary with the stripe embryo. The results of the
computer simulation are presented in Table 2. Analysis of the computer
simulation data for the different GBs shows, that GB $\Sigma5$, $\theta =
53.13 \, \mathrm{^{\circ}} $ with the smallest repetition distance has
negative $E_b^{GB+emb}$. All other analyzed GBs have positive $E_b^{GB+emb}$%
. It means that there is an energy advantage of the embryo stripe formation
in all of these GBs. There is some correlation between $E_{b}^{GB+emb}$ and $%
\Sigma$, but much more distinct dependence close to linear is found between $%
E_{b}^{GB+emb}$ and the repetition distance $R_{d}$ of the GB as shown in
Fig. 4. This dependence can be approximated by the best-fit linear
regression line $\ y=33.366x-89.112$, with the coefficient of determination $%
R^{2}=0.833$. Thus, there is a linear correlation between $E_{b}^{GB+emb}$
and repetition distance $R_{d}$. We think that this linear regression
approximation can be used to estimate the $E_{b}^{GB+emb}$ not only for GBs
analyzed in this article but for different GBs observed in physical
experiments.

It was shown above that there is energy disadvantage of the formation of
charge order by comparatively long chains of charged vacancies in the bulk.
At the same time, the stripe embryo formation at all high-angle GBs (with
the exclusion only GB $\Sigma5$, $\theta = 53.13 \, \mathrm{^{\circ}} $)
leads to decrease of energy and allows additional vacancies to get into
formation with the stripe embryo. One can expect that a type of GB may
determine the character of the initiated charge order inside of a grain, and
that GBs may play a special role in the adjustment of charge orders in
neighboring grains. This assumption, of course, should be strictly examined.
The presence of stripes may substantially influences the transport
properties of GBs in high-Tc superconductors. It is difficult now to
estimate this influence, but if the experimental studies of the charge order
in the high-Tc superconductors usually were performed in the single
crystals, one can expect that corresponding experiments performed on the
bi-crystals containing high-angle GBs also to be of interest to charge order
formation and its influence on the transport properties of GBs in the YBCO.

\section{Conclusion}

Thus, based on approach \cite%
{KhomskiiKugelEurophysLett2001,KhomskiiKugelPhysRevB2003} and our results,
one can conclude that there is a possibility of the charge order formation
consisted of the charged oxygen vacancies or di-vacancies in the YBCO. The
high-angle GBs regions are preferable places for the charge order formation
in the YBCO. It should be interesting to verify experimentally on the
bi-crystals containing the high-angle GBs the charge order formation and its
influence on the transport properties of the GBs in the YBCO. It can be
expected that obtained above results might be qualitatively applied to the
other cuprate high-temperature superconductors because they consist of the
parallel Cu-O planes, similar to such in the YBCO, with other elements
sandwiched in between these planes.

%-------------------------------------------------------------------------
%-------------------------------------------------------------------------

\textbf{Acknowledgment:} The authors express thanks to A.B. Kuklov for a
useful discussion. This work is supported by grant from the City University
of New York PSC-CUNY Research Award Program (Project Number: 68061-0037).

\end{document}